\documentclass{INTERSPEECH2023}

\interspeechcameraready

\newcommand\numberthis{\addtocounter{equation}{1}\tag{\theequation}}
\usepackage{amsmath,amssymb,graphicx,mathtools, nccmath}
\usepackage{caption,subcaption}
\usepackage{multirow}
\usepackage{float}
\usepackage{lipsum}
\usepackage{hyperref}
\usepackage{enumitem}
\usepackage[nodisplayskipstretch]{setspace}
\usepackage{booktabs}
\usepackage{mathtools}
\usepackage{etoolbox}
\usepackage{xcolor}
\usepackage[capitalize]{cleveref}
\usepackage{blindtext}
\usepackage{eso-pic,rotating,graphicx}
\usepackage{microtype}

\makeatletter
\renewcommand{\section}{\@startsection
	{section}%
	{1}%
	{}%
	{-0.7\baselineskip}%
	{0.1\baselineskip}%
	{}}%
\renewcommand{\subsection}{\@startsection
	{subsection}%
	{2}%
	{}%
	{-0.3\baselineskip}%
	{0.1\baselineskip}%
	{}}%
\renewcommand{\subsubsection}{\@startsection
	{subsubsection}%
	{3}%
	{}%
	{-0.1\baselineskip}%
	{0.1\baselineskip}%
	{}}%
\g@addto@macro\normalsize{%
	\setlength\abovedisplayskip{5pt plus 2pt minus 2pt}
	\setlength\belowdisplayskip{5pt plus 2pt minus 2pt}
	\setlength\abovedisplayshortskip{4pt plus 2pt minus 2pt}
	\setlength\belowdisplayshortskip{4pt plus 2pt minus 2pt}
}
\makeatother
\newenvironment{blurb}
{\par\scriptsize}
{\par\addvspace{\bigskipamount}}

\title{Competitive and Resource Efficient Factored Hybrid HMM Systems are Simpler Than You Think}
\name{Tina Raissi$^{*1}$\thanks{$^*$Denotes equal contribution}, Christoph L\"uscher$^{*1,2}$, Moritz Gunz$^{*1}$, Ralf Schl\"uter$^{1,2}$, Hermann Ney$^{1,2}$}

\address{
$^1$Machine Learning and Human Language Technology,\\
Computer Science Department, RWTH Aachen University, 52074 Aachen, Germany\\
$^2$AppTek GmbH, 52062 Aachen, Germany
}
\email{\{raissi,luescher,schlueter,ney\}@cs.rwth-aachen.de, moritz.gunz@rwth-aachen.de}

\begin{document}
	
	\maketitle
	
	\begin{abstract}
		Building competitive hybrid hidden Markov model~(HMM) systems for automatic speech recognition~(ASR) requires a complex multi-stage pipeline consisting of several training criteria.\ The recent sequence-to-sequence models offer the advantage of having simpler pipelines that can start from-scratch.\ We propose a purely neural based single-stage from-scratch pipeline for a context-dependent hybrid HMM that offers similar simplicity. We use an alignment from a full-sum trained zero-order posterior HMM with a BLSTM encoder.\ We show that with this alignment we can build a Conformer factored hybrid that performs even better than both a state-of-the-art classic hybrid and a factored hybrid trained with alignments taken from more complex Gaussian mixture based systems.\ Our finding is confirmed on Switchboard 300h and LibriSpeech 960h tasks with comparable results to other approaches in the literature, and by additionally relying on a responsible choice of available computational resources. 
		
	\end{abstract}

	\noindent\textbf{Index Terms}: automatic speech recognition, context-dependent acoustic modeling, hybrid hmm

	\section{Introduction}
	\label{intro}

	The first stage of a common pipeline for building a hidden Markov model~(HMM) based hybrid system includes bootstrapping of a context-dependent Gaussian mixture model~(GMM) with optional speaker adaptation.\ The phonetic context is usually modeled by allophones requiring state-tying via classification and regression trees~(CART)~\cite{odell1994tree}.\ These two building blocks serve the purpose of: (1) obtaining the alignment used during frame-wise cross-entropy training of the neural network, i.e. Viterbi training with fixed path, and (2) determination of the set of labels associated with the hidden states in the underlying Markov chain.\ This standard approach comprises of a well-known two-fold issue.\ The state-of-the-art hybrid systems rely generally on higher-level speech representations based on powerful encoder modules such as bidirectional long-short term memory~(BLSTM), Transformer~\cite{dong2018speech}, or Conformer~\cite{gulati2020conformer} whereas classic spectro-temporal or cepstral features are generally sufficient for GMMs.\ Therefore, the use of GMM alignments for building a competitive hybrid HMM system introduces an inconsistency due to the presence of different speech representations in different steps of the overall optimization process.\ In addition to this mismatch, there is also a more important inconsistency due to the inclusion of several training criteria in the overall pipeline.\ Other simplifying solutions such as use of lattice-free maximum mutual information~(LF-MMI) training criterion~\cite{endtoendpovey,zhang2021lattice} or factored hybrid HMM~\cite{raissi2020fh} can circumvent the mentioned issues by allowing from-scratch training and use of untied HMM states.\ However, sequence-discriminative training is generally more resource and time demanding than Viterbi training, with the ASR accuracy of the from-scratch LF-MMI being still questionable. Moreover, factored hybrid so far required at least a context-independent GMM alignment for obtaining comparable results to classic CART based systems.
	
	The current sequence-to-sequence~(seq-2-seq) models can be trained without a given alignment using sequence-level cross-entropy~(full-sum)~\cite{chorowski2015attention,ctc,rnnt,rnaSak}.\ However, despite sharing the same common denominator as an end-to-end system, the joint optimization of acoustic and internal language models pose important challenges for the combination of an external language model.\ Apart from different internal language model subtraction methods~\cite{zeineldeen2021investigating,meng2021internal}, an interesting case study focused on the acoustic model is the hybrid autoregressive transducer~\cite{variani2020hybrid}, and its recent extension~\cite{meng2022modular}. These methods aim at solving an issue that comes along with the direct discriminative approach in the first place.\ The classic hybrid formulation on the other hand offers a clear distinction of the acoustic and language model, intrinsically.\
	
	In this work, we leverage the potential of the factored hybrid model for a from-scratch pipeline that does not make use of any of the standard pipeline components such as Gaussian models or clustering algorithms. We show that by using the alignment from a full-sum trained BLSTM posterior HMM it is possible to build a single-stage training pipeline with comparable results to other approaches in the literature.\ We also show an additional robustness to the choice of n-gram phonetic context when using a Conformer encoder.\ Our experiments show competitive results on both Switchboard~(SWB) 300h and LibriSpeech~(LBS) 960h tasks.\ We also limited the use of computational resources by (1) carrying out a single training run on the SWB task, only with the findings from the LBS task, and (2) tuning the acoustic and language model-related scales for the time-synchronous beam search with use of an additional pruning method based on acoustic lookahead.
	
	\section{Modeling Approach}
	\label{approch}	
	For an acoustic feature sequence $X$ and a word sequence $W$, define $h_1^T = E(X)$ to be the encoder output sequence of length $T$ that transforms $X$ to higher level representations.\ We consider the output sequence to be a phoneme sequence $a_1^S$ corresponding to $W$.\ For the underlying hidden state sequence $s_1^T$, an allophone state label identity is associated to each hidden state.\ Each allophone consists of single-state left and right phonemes and a tripartite center phoneme with end-of-word augmentation.\ Define $a_{s_t}$ to be the aligned phoneme state at time frame $t$. For simplicity, we denote the single-state left and right phonemes with $a_{s_t}{-}{1} = a_{s_t}^{\ell}$ and $a_{s_t}{+}{1} = a_{s_t}^{r}$, respectively.\ The allophone state set at each time step can be then defined as $\{a_{s_t}^{\ell},a_{s_t}, a_{s_t}^{r}\}$.\ The general modeling approach for a hybrid HMM defines:
	
	\begin{align*}
		P(h_1^T | a_1^S)& = \underset{s_1^T}{\sum} P(h_1^T, s_1^T | a_1^S) \\
		&= \underset{s_1^T:a_1^S}{\sum} \prod_{t=1}^T P(h_t |a_{s_t}^{\ell},a_{s_t}, a_{s_t}^{r}) P(s_t | s_{t-1}) \\
		&= \underset{s_1^T:a_1^S}{\sum} \prod_{t=1}^T \frac{P(a_{s_t}^{\ell},a_{s_t}, a_{s_t}^{r} | h_t)}{P_{\text{prior}}(a_{s_t}^{\ell},a_{s_t}, a_{s_t}^{r})}P(s_t | s_{t-1}) \numberthis \label{eq:hmm}
	\end{align*}
	We follow the forward factorization of the joint label posterior probability in \cref{eq:hmm} as done in~\cite{raissi2020fh}: 
	
	\begin{align*} 
		\hspace{-0.2cm} P(a_{s_t}^{\ell},a_{s_t}, a_{s_t}^{r} | h_t){=}P(a_{s_t}^{\ell}| h_t)P(a_{s_t} |  a_{s_t}^{\ell}, h_t) P(a_{s_t}^{r} |a_{s_t}, a_{s_t}^{\ell}, h_t )
	\end{align*}	
	The quantity $P(a_{s_t} |  a_{s_t}^{\ell}, h_t)$ is similar to the label posterior at time frame $t$ in a (phoneme-) transducer with a first-order context dependency, i.e. a diphone transducer.\ However, in addition to the blank-augmented frame label $y_t=a_{s_t}$, the transducer starts by directly modeling the $P(a_1^S | h_1^T )$ and therefore does not need the additional monophone probability for the left context.\ Moreover, in factored hybrid by relying on a generative starting point one can include the right context as part of the model definition, since the right context target appears as a dependency and not as a direct prediction.\ Moreover, the use of the context-dependent priors for each factor serves as a local normalization.\ 
	
	During decoding, we use the log-linear combination of acoustic and language model, as defined in~\cite{raissi2022fh}.

	\section{Effective Viterbi Training without GMM}
	
	The use of sequence level training criteria that sum over all paths in the lattice of the possible alignments allows for from-scratch training of not only the common seq-to-seq models but also 0-order neural HMMs.\ However, a careful choice of the label topology and unit, as well as the corresponding input frame rate is necessary in order to counteract convergence problems during training.\ The frame-wise cross-entropy training with a given alignment, i.e. Viterbi training, can reduce the required number of optimization steps due to a faster convergence.\ The Viterbi training can also take advantage of several additional regularization techniques, such as division of the sequence into chunks and use of auxiliary and complementary losses.\ We used the explicit phoneme labels both with and without context-dependency as targets for additional losses during training.\ We show the contribution of such losses across different encoder architectures and models.\ Furthermore, we examined the performance of our factored hybrid models when using alignments from systems of different complexities and compared this with the standard GMM/CART pipeline.\

	\begin{table}[t]
	
	\setlength{\tabcolsep}{0.4em}\renewcommand{\arraystretch}{1.}  
	\centering \footnotesize 
	\caption{Inclusion of additional tasks on left, center or right phonemes for CART and triphone factored (FH), using SWB 300h and LBS 960h.\ All models use context-dependent GMM alignments described in \cref{sub:align} .\ In case of FH the main loss is on center phoneme state.\ }
	\label{tab:multi}
	\begin{tabular}{|c|c||c|c|c||c|c|} 
		\hline 
		\multirow{2}{*}{\textbf{Enc.}} &  \multirow{2}{*}{ \textbf{Model} }  &\multicolumn{3}{c||}{ \textbf{Multi-Task} } & \multicolumn{2}{c|}{ \textbf{WER [\%]}} \\ \cline{3-7}
		& 	 & \textbf{left} &  \textbf{center} &\textbf{right} &  {Hub5'00}& {dev-other} \\ \hline
			\multirow{4}{*}{Confo}& \multirow{2}{*}{CART} &  \multicolumn{3}{c||}{no} &$11.6$&$7.4$ \\ \cline{3-7} 
		&& \multicolumn{3}{c||}{yes}&$11.4$& $7.2$\\ \cline{2-7}
		&Triphone &  \multicolumn{3}{c||}{no}&$11.6$ & $7.1$\\ \cline{3-7}
		& FH& no&yes &no &$11.5$&$7.0$ \\ \hline \hline
		\multirow{6}{*}{BLSTM}&\multirow{2}{*}{Monophone} &\multicolumn{3}{c||}{no} & $14.3$&  \multirow{6}{*}{ - }  \\ \cline{3-6} 
		& & yes& no &yes &$13.6$&  \\ \cline{2-6} 
		& \multirow{2}{*}{CART} &  \multicolumn{3}{c||}{no} &$12.9$&\\ \cline{3-6} 
		&& \multicolumn{3}{c||}{yes} &$12.9$& \\ \cline{2-6}
		&Triphone &  \multicolumn{3}{c||}{no} & $12.7$& \\ \cline{3-6}
		&FH& no& yes&no&$12.6$& \\ \hline 
				
	\end{tabular} 

\end{table}
		\begin{table}[t]
	\setlength{\tabcolsep}{0.8em}\renewcommand{\arraystretch}{1.}  
	\centering \footnotesize
	\caption{Performance of Conformer monophone and factored hybrid models with varying auxiliary losses, evaluated on LBS dev-other using a 4-gram LM.}
	\label{tab:LBS-aux}
	\begin{tabular}{|c||c|c|c|}
		\hline
		\multirow{2}{*}{ \textbf{Aux. Loss} } & \multicolumn{3}{c|}{ \textbf{dev-other [\%]}} \\ \cline{2-4}
		& mono & di & tri \\ \hline
		Baseline & $6.9$ & $6.8$ & $6.7$ \\ \hline
		Best combination & $6.9$ & $6.6$ & $6.6$ \\
		\hline
	\end{tabular}
\vspace{-0.2cm}
\end{table}

	\subsection{Phonetic Multi-Task Criteria and Auxiliary Losses} 
	\label{sub:multi}
	The classic hybrid HMM realizes the acoustic-phonetic context-dependency via allophone state targets, that are clustered during training to avoid sparsity issues.\ However, the use of secondary tasks that consider additional label targets of different granularity have been shown to report improvements~\cite{bell2016multitask,raissi2020fh}.\ The distinctive property of an HMM label topology is the frame-wise label emission that can be restrained to mere loop and forward transitions.\ We take advantage of this commonly known 0-1 HMM topology for further regularizing the training.\ At each time frame, we consider the label of the neighboring right and left phonemes of the aligned allophone state and use them as additional tasks and auxiliary losses.\ This is different from the blank-based models.\ The introduction of the blank symbol in CTC or transducers relaxes the classic HMM topology and the associated transition model.\ The model has more degrees of freedom at each input step: from emitting more than one label at each step in case of classic transducer to the possibility of emitting several blank labels for consecutive time frames.\ This leads to arbitrary alignments that are not suitable for such regularization techniques.\ 
	
	The factored hybrid model in this case has the additional advantage of consisting of several output targets, by definition.\ In contrast to the CART model which has only one output, an $n$-gram factored model requires the combination of $n$ probability distributions for the calculation of the final distribution over the whole set of untied triphone states, which is in the order of $| phonemes |^3$.\ Since  in practice the application of the time-synchronous beam search for models with $n$-gram order higher than three is not feasible, we fix this number as our upper bound.\ Within this limit, one can make use of additional context-dependent or context-independent label outputs on different levels of the neural architecture. We distinguish between two different cases:
	
	\begin{itemize}[leftmargin=*, itemsep=-0.5mm]
		\item \textbf{Multi-task loss}: used for the back-end model including the output targets that are not used during decoding.
		\item \textbf{Auxiliary loss}: applied on the encoder intermediate layers.
	\end{itemize}
Our comparison is done on two different levels.\ We show the impact of multi-task losses on CART models for both Conformer and BLSTM encoders.\ We then examine a factored triphone model with an additional loss on only the center phoneme, where all six states, i.e. the three HMM states and the respective word-end class distinctions, are tied together.\ This is shown in \cref{tab:multi}.\ It is possible to see that CART model with BLSTM encoder does not take advantage of the three additional tasks.\ However, this is different in a Conformer model.\ The factored triphone model had slightly better performance.\ This behavior is consistent also on the LBS task for the Conformer model.\ Finally, the largest difference is seen in a BLSTM monophone model on SWB.\ However, for Conformer monophone models we did not observe noticeable difference.\ Therefore we looked into different auxiliary losses.\ For a 12-layers Conformer architecture, we selected different candidate layers on which the auxiliary losses have been applied.\ We considered one center phoneme loss at layer 6 as the baseline.\ The additional losses on all three left, center, and right phonemes were positioned at layers $\mathcal{L}_1 = \{3,6,9\}$ or $\mathcal{L}_2 = \{4,8\}$, with a context order that was increasing either bottom-up or top-down.\ We found that monophone model again did not take advantage of the auxiliary losses.\ However, for the layer set $\mathcal{L}_1$ diphone and triphone models could obtain the improvements shown in \cref{tab:LBS-aux} with bottom-up and top-down increasing context-dependency order, respectively.

\begin{table}[t]		
	\setlength{\tabcolsep}{0.8em}\renewcommand{\arraystretch}{1.}  
	\centering \footnotesize 
	\caption{Performance of Conformer factored hybrid~(FH) and CART models trained using BLSTM Posterior HMM~(P-HMM), GMM monophone~(GMM-Mono), and Tandem alignments, evaluated on HUB5'00 with a 4gram LM.}
	\label{tab:SWB-align}
	\begin{tabular}{|c|c||c|c|} 
		\hline 
		\multirow{2}{*}{ \textbf{Alignment} } &  \multirow{2}{*}{ \textbf{Model} } &\multicolumn{2}{c|}{ \textbf{HUB5'00 [\%]}} \\ \cline{3-4}
		& 	 &      di & tri \\ \hline
		\multirow{2}{*}{TANDEM}  & CART &  \multirow{2}{*}{-} &$11.6$ \\ \cline{2-2} \cline{4-4}
		& \multirow{3}{*}{FH} &  & $11.6$ \\ \cline{1-1} \cline{3-4}
		GMM-Mono& FH &  $12.4$ & $12.4$  \\ \cline{1-1} \cline{3-4}		
		P-HMM&   &  $11.5$ & $11.4$ \\ \hline
		
	\end{tabular} 
	
\end{table}

\begin{table}[t]
	
	\setlength{\tabcolsep}{0.8em}\renewcommand{\arraystretch}{1.}  
	\centering \footnotesize 
	\caption{Results on LBS task with the most promising alignments introduced in \cref{tab:SWB-align}, and evaluations done on dev-other using a 4-gram LM.\ }
	\label{tab:LBS-align}
	\begin{tabular}{|c|c||c|c|} 
		\hline 
		\multirow{2}{*}{ \textbf{Alignment} } &  \multirow{2}{*}{ \textbf{Model} } &\multicolumn{2}{c|}{ \textbf{dev-other [\%]}} \\ \cline{3-4}
		& 	 &      di & tri \\ \hline
		\multirow{2}{*}{GMM-Tri}  & CART & $7.6$ & $7.4$ \\ \cline{2-4}
		& \multirow{2}{*}{FH} &  $7.1$ & $7.1$ \\ \cline{1-1} \cline{3-4}
		P-HMM&  &$7.1$ & $7.0$ \\ \hline
	\end{tabular}
	
\end{table}
\begin{table}[t]
	\setlength{\tabcolsep}{0.03em}\renewcommand{\arraystretch}{1.}  
	\centering \footnotesize 
	\caption{Comparison of different encoder architectures using GMM/CART based alignments against neural monophone P-HMM alignment.\ The GMM alignments are the Tandem and GMM-Tri, described in \cref{sub:align}, for SWB 300h and LBS 960h, respectively.\ Decoding is done by using 4-gram LM. }
	\label{tab:baselines}
	\begin{tabular}{|c|c|c|c||c|c||c|c|} 
		\hline 
		\multirow{2}{*}{\textbf{Enc.}} & \multirow{2}{*}{ \textbf{Align.} } &  \multirow{2}{*}{ \textbf{Model} } & \multirow{1}{*}{ \textbf{n-} } &\multicolumn{4}{c|}{ \textbf{WER [\%]}} \\ \cline{5-8}
		& 	 &    & \textbf{phone} &  {Hub5'00}& {Hub5'01} & dev-other & test-other \\ \hline \hline
		\multirow{3}{*}{Confo}  & GMM& CART &  \multirow{2}{*}{3}&$11.6$&$11.4$&$7.6$& $7.7$\\ \cline{2-3} \cline{5-8}
		& \multirow{2}{*}{P-HMM} & \multirow{2}{*}{FH} &&$11.4$& $11.4$&$7.0$&$7.3$ \\ \cline{4-8}
		&&&2 &$11.5$&$11.0$&$7.1$& $7.4$\\ \hline
		\multirow{2}{*}{BLSTM}  & GMM &CART & \multirow{2}{*}{3}&$12.9$&$13.2$& \multicolumn{2}{c|}{ \multirow{3}{*}{-} }\\  \cline{2-3}  \cline{5-6}
		& \multirow{2}{*}{P-HMM} &  \multirow{2}{*}{FH}&&$12.7$&$12.9$& \multicolumn{2}{c|}{ }  \\ \cline{4-6}
		&&&2&$13.8$&$13.6$&\multicolumn{2}{c|}{ }  \\ \hline
		
	\end{tabular} 
	\vspace{-0.4cm}
\end{table}

	\subsection{Effect of the Alignment Model}
	\label{sub:align}
	
	The finite state acceptor~(FSA) structure used for the alignment model is a standard HCLG~\cite{mohri2004weighted}.\ It is possible to decouple the n-gram order of the context-dependency used for the construction of the graph (the C in HCLG) from the alignment model.\ This means that during the construction of the alignment FSA we can consider all possible paths resulting from the allophone sequence corresponding to a word sequence, by including left and right context for each phoneme.\ However, one could apply a different level of state-tying for the alignment model, e.g. a monophone state-tying.\ On the other hand, due to the explicit context-dependency modeled via the phoneme embeddings in factored hybrid, it is possible to use the alignment taken from a monophone model and train a diphone or triphone factored models.\ We considered four different alignments for SWB and LBS tasks, as follows:
	\begin{itemize}[leftmargin=*, itemsep=-0.5mm]
		\item \textbf{P-HMM}: monophone single-stage BLSTM Posterior HMM trained with sequence-level cross-entropy from-scratch~\cite{raissi2022hmm}
		\item \textbf{GMM-Mono}: monophone GMM alignment
		\item  \textbf{GMM-Tri}: triphone GMM alignment for LBS~\cite{luscher2019rwth}
		\item \textbf{TANDEM}: tandem based triphone of a complex multi-stage NN-GMM pipeline for SWB~\cite{tuske2015asru}
	\end{itemize}
	The first alignment uses a discriminative HMM with no GMM involved.\ The last two alignments use state-tying via CART.\ All alignment models are trained on the same FSA structure.\ In \cref{tab:SWB-align,tab:LBS-align}, we show the effect of the choice of the alignment model on the ASR accuracy of the Conformer factored hybrid and compared this against a CART model trained on GMM-Tri and TANDEM alignments for LBS and SWB tasks, respectively.\ All ASR models are built using the same training parameters and number of epochs, described in \cref{exp}.\ On both tasks, factored hybrid can reach better results when using P-HMM alignment, by avoiding the classic complex pipeline with heterogeneous optimization criteria.\ On SWB task, we observed up to $9\%$ relative degradation when using the GMM-Mono alignment.\ We excluded this experiment on LBS, accordingly.

\subsection{Elimination of GMM and State-Tying}
	\label{sub:expsingle}
	We selected the best auxiliary loss variation from \cref{sub:multi}, by doing majority voting between the three mono-/di-/triphone models.\ This resulted to be a bottom-up approach using layer set $\mathcal{L}_1$ described in \cref{sub:multi}.\ This was the best strategy according to monophone and diphone experiments.\ We then built diphone and triphone models using the P-HMM alignment on both tasks.\ We compare the single-stage Viterbi-trained factored models against the triphone CART system using the most complex alignment model.\ Our results in \cref{tab:baselines} show that on both tasks, regardless of the encoder architecture, factored hybrid can reach the same performance compared to CART and in some cases can outperform it.\ For the Conformer model on SWB, the diphone model generally performed better.\ We think this might be due to the choice of the auxiliary loss that was optimal for the diphone model.\

	\section{Experimental Details}
	\label{exp}
	
    The experiments are conducted on 300h Switchboard-1 (SWB) Release 2 (LDC97S62) \cite{godfrey1992SWB} and 960h LibriSpeech (LBS) \cite{povey2015librispeech}.
    The evaluations for the SWB task are performed on SWB and CallHome subsets of Hub5`00 (LDC2002S09) and three subsets of Hub5`01 (LDC2002S13).
    For LBS we report WERs on dev and test sets.
    \ifinterspeechfinal
    We utilize the toolkits RETURNN \cite{zeyer2018:returnn} and RASR~\cite{wiesler2014rasr}.
    \else
    We utilize two toolkits for training and decoding \cite{zeyer2018:returnnblind,wiesler2014rasrblind}.
    \fi
    All models are trained with frame-wise cross entropy criterion with an external alignment.
    The alignments are described and referenced in \cref{sub:align}. 
    The speech signal is represented by (SWB:~40, LBS:~50) dimensional Gammatone filterbank features, extracted from a $25$ms window with a $10$ms shift~\cite{schluter2007Gammatone}.
    The state inventory of both corpora consists of the complete set of triphone states, corresponding to a tripartite 0-1 HMM label topology for each phoneme in context, and a special context-independent silence state. Differently to the standard LBS phoneme inventory, we removed stress markers from all phonemes. For the standard hybrid, a set of (SWB:~9001, LBS:~12001) CART labels are considered.
    The learning rate~(LR) schedule is either fixed with one cycle learning rate or we apply Adam optimizer with Nestorov momentum~\cite{dozat2016incorporating}.
    We use gradient noise of (SWB:~$0.1$, LBS:~$0.0$) and an optimizer epsilon of $1e^{-08}$ for all experiments.
    We set a minimum LR of (SWB:~$2e^{-05}$, LBS:~$1e^{-06}$).
    SpecAugment is applied to all models.\\
%
%
    Our Conformer \cite{gulati2020conformer} encoder follows the setups from \cite{luescher2023embedding} and is employed for experiments on SWB and LBS.
    The Conformer models train for (SWB:~50, LBS:~15) epochs. 
    After the Gammatone feature extraction and the application of SpecAugment, three frames are stacked and a time downsampling factor of 3 is applied via a combination of convolutional and pooling layers.
    The intermediate representations are then passed on to 12 Conformer blocks, followed by upsampling of the encoder output to the original time length.
    The model size of our Conformer is between 80-87M parameters, depending on the model being CART or FH.
    We apply label smoothing $0.2$ only for FH models following~\cite{raissi2022fh} and a focal loss factor $2.0$ to the loss calculation.
    For our Conformer training we apply a one cycle learning rate schedule (OCLR) \cite{smith2019super}.
    We used OCLR with a peak LR of around $1e^{-3}$ over 90\% of the training epochs, followed by a linear decrease to $1e^{-6}$.\\ 
    %
    %
    We utilize a recurrent encoder in our SWB experiments and keep the hyperparameters for the CART and FH setups identical,
    following a similar setup as in \cite{raissi2022fh}.
    The recurrent encoder consists of 6 BLSTM layers with 512 nodes per direction, resulting in $\sim$46M model parameters.
    We apply dropout to each LSTM layer with a probability of $10\%$.
    The models train for 50 epochs with a batch size of 10k.
    We chunk each sequence into 64 frames and a shift of 32 frames.
    The learning rate schedule follows the Conformer recipe described above.
    For recognition we use the official 4-gram language model and a (SWB:~LSTM, LBS:~Transformer) language model~\cite{beck2019lstm,beck2020lvcsr}.
    Both models utilize a lexical prefix tree and a time-synchronous beam decoding with dynamic programming for search space organisation.
    For decoding we tune the LM, prior, and time distortion penalty~(tdp) scales, together with the lemma-level silence exit penalty.\ This is done via an efficient two-stage grid search.\ We first set a very small beam together with a high scale for acoustic look-ahead with temporal approximation~\cite{nolden2013advanced} and tune all mentioned values except for the LM scale.\ Subsequently, we set our final beam and tune the LM scale by performing rescoring on the resulting lattices.\ We then run the final decoding with the optimum LM scale and report word error rates.
    For further details on training hyper parameters and decoding settings, we refer to an example our configuration setups\footnote{\ifinterspeechfinal \scriptsize{ \url{ https://github.com/rwth-i6/returnn-experiments}}\else BLIND\fi}.

    \begin{table}[t]
    	\setlength{\tabcolsep}{0.3em}\renewcommand{\arraystretch}{1.}  
    	\centering \footnotesize
    	\caption{Requirements in terms of training time and hardware resources for the GMM and posterior HMM alignment models for LBS 960h.\ }
    	\label{tab:alignment}
    	\begin{tabular}{|c||c|c|c|}
    		\hline
    		\multirow{2}{*}{ \textbf{Alignment model} } & \multicolumn{2}{c|}{ \textbf{Resources}} & \multirow{2}{*}{ \textbf{ \# Stages} }\\ \cline{2-3}
    		& \# Hours & Hardware &  \\ \hline
    		GMM & $6376$ & $1$ CPU & $6$ \\ \hline
            BLSTM Posterior HMM & ~~$417$ & $1$ GPU & $1$ \\
    		\hline
    	\end{tabular}
    	\vspace{-0.4cm}
    \end{table}

	\subsection {Comparison with Literature}
    Standard hybrid systems take advantage of speaker adaptation and sequence discriminative training in order to reach competitive results.\ In our proposed work, we examined only the Viterbi training stage results.\ Therefore, we compare the performance of our from-scratch factored hybrid models with only single-stage approaches in the literature.\ Regarding the computational resources and training time in terms of hours for the alignment model, it is important to note that the standard hybrid model usually requires a multi-stage pipeline for a context-dependent GMM alignment with speaker adaptation.\ A comparison in terms of hours of training and necessary hardware resources is reported in \cref{tab:alignment}.\ Our monophone BLSTM Posterior HMM alignment model is trained on an outdated NVIDIA GTX 1080 Ti GPU, leaving the training speed improvement a viable option when switching to a more modern GPU.\ The GMM alignment model training on the other hand allows for parallel computation and a significant reduction of the training hours given a sufficiently large CPU cluster.\ Therefore, the computational resources required for both alignment models differs starkly, the GMM alignment model requiring a full cluster for a speedy generation while the Posterior HMM alignment relying on a single machine.\ Regarding the ASR accuracy, it is possible to see that the factored hybrid with our proposed simplified pipeline can reach competitive results.\ For SWB task, as shown in \cref{tab:SWB-lit}, the factored model outperforms all other approaches on HUB5`01, including the attention encoder-decoder~(AED) model consisting of many more parameters and trained for many more epochs.\ Similar scenario is valid also for the results on LBS, with the exception of the E-Branchformer~(E-Brch).\ However, it is important to note that in addition to larger number of epochs, the model combination during training and decoding, as well as model checkpoint averaging do not allow for a fair comparison to our proposed approach.\ Furthermore, our model offers more flexibility when switching domain due to the separate language model.\
	
	    \begin{table}[t]
		
		\setlength{\tabcolsep}{0.2em}\renewcommand{\arraystretch}{1.1}  
		\centering \footnotesize 
		\caption{Results on averaged dev and test for SWB 300h for single-stage pipelines of different models using 4-gram and LSTM LMs.}
		\label{tab:SWB-lit}
		\begin{tabular}{|c||c|r|c|c||c|c|} 
			\hline 
			\multirow{2}{*}{ \textbf{Work} } & \multicolumn{3}{c|}{ \multirow{1}{*}{ \textbf{Model}}}& \multirow{2}{*}{\textbf{LM }}&\multirow{2}{*}{\textbf{HUB5'00}} & \multirow{2}{*}{\textbf{HUB5'01}}\\ \cline{2-4}
			&Topology & \#PMs& \#EPs&  & &\\ \hline \hline
			\cite{zeineldeen2022improving} & Confo Hybrid & 58M & ~~50 & \multirow{2}{*}{4-gram}&$10.7$ & $11.0$\\ \cline{1-4} \cline{6-7}
			\cite{weiefficient} & Confo Trans &75M & ~~50 &  & $11.4$ & - \\\cline{1-7}
			\cite{zoltan;AED} & RNN AED & 280M & 250 & LSTM &$\mathbf{9.8}$ &$10.1$ \\ \hline \cline{1-7}
			\multirow{2}{*}{This} &  \multirow{1}{*}{Confo FH} &  \multirow{2}{*}{80M} &  \multirow{2}{*}{~~50} & 4-gram & $10.7$ & $10.6$ \\ \cline{5-7}
			& (P-HMM) &  &   &   LSTM &$10.0$ & $\mathbf{9.9}$ \\ \hline
			
		\end{tabular} 
	\end{table}
	
	\begin{table}[t]

		\setlength{\tabcolsep}{0.07em}\renewcommand{\arraystretch}{1.1}  
		\centering \footnotesize 
		\caption{Results on all four dev and test sets for LBS 960h for single-stage pipelines of different models using 4-gram, LSTM, and Transformer LMs.}
		\label{tab:LBS-lit}
		\begin{tabular}{|c||c|r|c|c||c|c|c|c|} 
			\hline 
			\multirow{2}{*}{ \textbf{Work} } & \multicolumn{3}{c|}{ \multirow{1}{*}{ \textbf{Model}}}& \multirow{2}{*}{\textbf{LM }}& \multicolumn{2}{c|}{ \multirow{1}{*}{\textbf{dev}}} &  \multicolumn{2}{c|}{ \multirow{1}{*}{\textbf{test}}} \\ \cline{2-4} \cline{6-9}
			& Topology & \#PMs& \#EPs&  &clean & other &clean & other\\ \hline \hline
			\cite{weiefficient} & Confo Trans & 75M & ~~20 & 4-gram& $2.9$ & $6.9$&  \multicolumn{2}{c|}{-}  \\ \hline
			\cite{park2019specaugment} & RNN AED & 360M & 600 & LSTM &  \multicolumn{2}{c|}{ \multirow{3}{*}{-}} &$2.2$ & $5.2$ \\ \cline{1-5} \cline{8-9}
			\cite{transformer2020Wang} & Trafo Hybrid &81M & 100 & \multirow{2}{*}{Trafo} &  \multicolumn{2}{c|}{} & $2.3$ & $4.9$ \\ \cline{1-4} \cline{8-9}
			\cite{kim2022ebranch} & E-Brch AED+CTC& 149M & ~~80 &  &  \multicolumn{2}{c|}{} &  $\mathbf{1.8}$ & $\mathbf{3.6}$ \\ \hline
			\multirow{2}{*}{This} &  \multirow{1}{*}{Confo FH} &  \multirow{2}{*}{80M} &  \multirow{2}{*}{~~15} & 4-gram & $2.9$ & $6.5$ & $3.3$ & $7.1$ \\\cline{5-9}
			& (P-HMM) &  &   &    Trafo &$2.1$ & $4.4$ & $2.3$ & $5.0$ \\ \hline		
		\end{tabular} 
		\vspace{-0.4cm}
	\end{table}
	
	\section{Conclusions}
	In this work we offered a simple from-scratch pipeline for a context-dependent hybrid HMM system. We avoided the common state-tying and GMM building blocks, by utilizing the alignment of a full-sum trained BLSTM posterior HMM and leveraging the explicit context modeling in factored hybrid.\ We observed that such alignment improves the accuracy of the subsequent Viterbi trained model, compared to GMM based alignment.\ Moreover, we showed that the phonetic context is learned differently when using BLSTM or Conformer encoders.\ A selection of examined different loss variants using phoneme targets was also discussed in this work.\ With our single-stage Viterbi trained models we obtained competitive results on both Switchboard 300h and LibriSpeech 960h. 
	
	\section{Acknowledgements}
	\begin{blurb}
		This work was partially supported by NeuroSys which, as part of the initiative “Clusters4Future”, is funded by the Federal Ministry of Education and Research BMBF (03ZU1106DA).\ This work was partially supported by a Google Focused Award.\ This work was partially supported by the project HYKIST funded by the German Federal Ministry of Health on the basis of a decision of the German Federal Parliament (Bundestag) under funding ID ZMVI1-2520DAT04A.\ The work reflects only the authors' views and none of the funding parties is responsible for any use that may be made of the information it contains.\ We thank Daniel Mann for running the alignment experiment for LibriSpeech task and Eugen Beck for valuable feedback and constant encouragement for the factored hybrid concept since its very beginning.
	\end{blurb}

	\bibliographystyle{IEEEtran}
	\bibliography{mybib}
	
\end{document}